\documentclass[conference]{IEEEtran}
\IEEEoverridecommandlockouts

\ifCLASSOPTIONcompsoc
  \usepackage[nocompress]{cite}
\else
  \usepackage{cite}
\fi

\usepackage{algorithm2e}
\RestyleAlgo{ruled}
\newcommand{\figref}[1]{Fig.~\ref{fig:#1}}
\usepackage{tabularx}
\usepackage{amsmath,amssymb,amsfonts}

\usepackage{booktabs} 

\usepackage{algorithmic}
\usepackage{graphicx}
\usepackage{textcomp}
\usepackage{xcolor}
\def\BibTeX{{\rm B\kern-.05em{\sc i\kern-.025em b}\kern-.08em
    T\kern-.1667em\lower.7ex\hbox{E}\kern-.125emX}}
\begin{document}

\title{ADARP: A Multi Modal Dataset for Stress and Alcohol Relapse Quantification in Real Life Setting
\thanks{$^{1}$ School of Electrical Engineering and Computer Science, $^{2}$ Elson S. Floyd College of Medicine, $^{3}$ Department of Human Development, $^{4}$ College of Health Solutions, $^{*}$ Washington State University, Pullman, USA, and $^{\#}$ Arizona State University, Tempe, USA. \\ Identify applicable funding agency here. If none, delete this.}
}

\author{\IEEEauthorblockN{Ramesh Kumar Sah$^{1, *}$, Michael McDonell$^{2, *}$, Patricia Pendry$^{3, *}$, Sara Parent$^{2, *}$, \\ Hassan Ghasemzadeh$^{4, \#}$, Michael J Cleveland$^{3, *}$}}

\maketitle

\begin{abstract}
Stress detection and classification from wearable sensor data is an emerging area of research with significant implications for individuals' physical and mental health. In this work, we introduce a new dataset, {\it ADARP}, which contains physiological data and self-report outcomes collected in real-world ambulatory settings involving individuals diagnosed with alcohol use disorders. We describe the user study, present details of the dataset, establish the significant correlation between physiological data and self-reported outcomes, demonstrate stress classification, and make our dataset public to facilitate research.
\end{abstract}


\section{Introduction}


Stress describes bodily reactions to perceived physical or psychological threats that trigger a cascade of physiological responses \cite{kasl}. Emotional responses to stress include anxiety, restlessness, and nervousness. Physiological responses to stress are governed by two interrelated systems, the hypothalamus-pituitary-adrenal (HPA) axis and the autonomic nervous system (ANS). The ANS affects different bodily functions depending on a person's emotional state. For example, it increases heart rate in the event of fight-or-flight response and reduces the heart rate to bring the body back to a rest state. Another bodily change induced by acute stress is sweat secretion. Sweating can be assessed by measuring the skin's electrical conductivity shift and is called Electrodermal Activity (EDA). EDA is particularly advantageous for stress monitoring because the skin is exclusively innervated by the sympathetic branch of the ANS, whereas most other organs are under the influence of both autonomic branches \cite{critchley2002electrodermal}.

\begin{figure}[!tbh]
    \centering
    \includegraphics[width=1.0\linewidth]{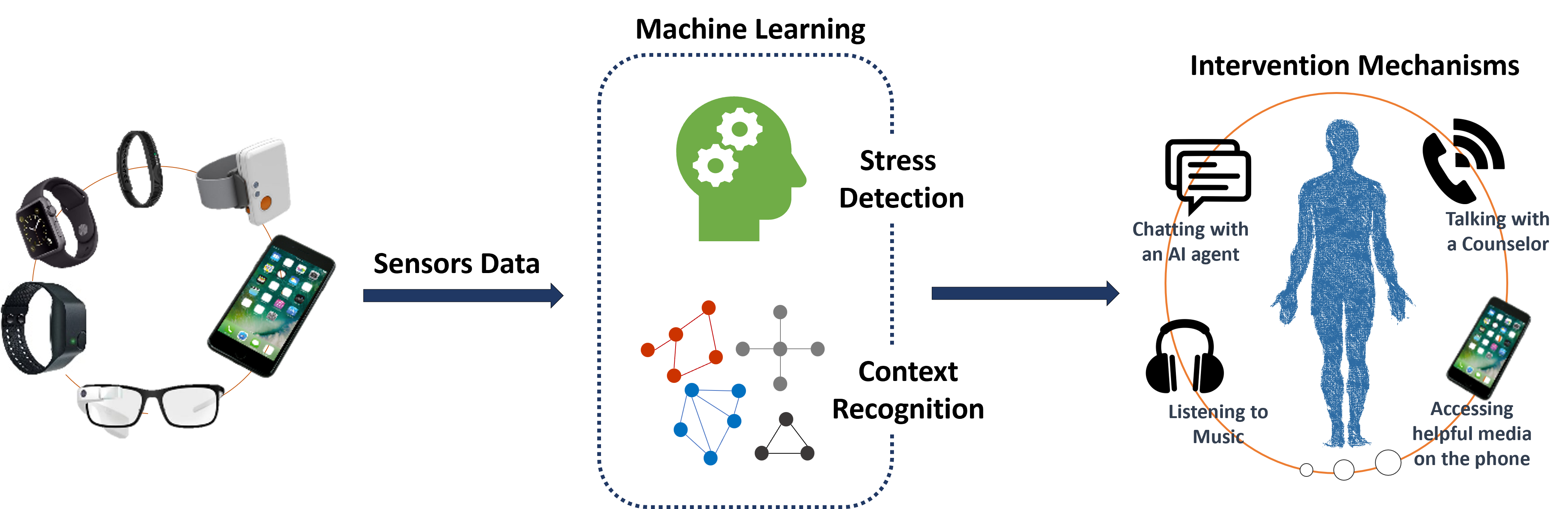}
    \caption{Stress detection, context recognition, and personalized intervention using wearable sensors and machine learning.}
    \label{fig:system}
\end{figure}

As shown in \figref{system}, wearable technologies offer considerable promise for stress monitoring and delivery of addiction treatments, including the use of ecological momentary assessment (EMA) as a data collection tool in which people report their moods and behaviors in real-time on mobile devices. Although EMA studies often include self-reports of perceived stress, there are many shortcomings to this assessment. Most importantly, self-reports only capture subjective aspects of stress and may not provide a full measure of the physical and health consequences of stress. Thus, biological markers of stress are commonly used to provide more proximal measures of physiological processes that react to physical and psychological demands. 

To date, most research has used controlled studies in laboratory settings to detect stress via physiological signals. However, induced stress in artificial laboratory settings may not fully represent stress that is experienced by individuals in their daily lives. In response, a nascent body of research has used ambulatory assessment of stress via wearable sensor technology to provide continuous, ambulatory monitoring of stress in uncontrolled, real-world settings \cite{smets2018into}. Most of this research has been conducted among healthy populations; thus, critical questions among clinical populations such as individuals diagnosed with substance use disorders remain unanswered. To the best of our knowledge, our work is the first to measure the effects of stress on alcohol addiction and relapse in a real-world setting using sensor systems and EMA surveys. In this work, we make the dataset public and share preliminary findings with the community to motivate and advance the frontiers of stress and addiction research. 

\section{Overview of ADARP Dataset}

\subsection{Study Protocol}
Our study, referred to as the {\it Alcohol and Drug Abuse Research Program} (ADARP) study, was a proof-of-concept pilot study aimed to discover how the daily experiences of patients diagnosed with alcohol use disorder (AUD) correspond with physiological biomarkers of stress. Each participant completed three components: 1) a daily diary study using ecological momentary assessment (EMA) of self-reported emotions, cravings, and stress via a web-based survey, prompted $4$ times daily for up to $14$ days; 2) continuous monitoring of stress with an Empatica E4 wristband that captured, in real-time, continuous physiological markers of stress, including heart rate (HR), skin conductance or electrodermal activity (EDA), skin temperature, and bodily movements; and 3) structured qualitative interviews to assess daily alcohol use, using a timeline follow-back calendar, and to validate self-reported and physiological markers of stress. 

We recruited a convenience sample of $11$ participants ($10$ female) from adult patients receiving treatment for mental health and AUD at a treatment agency in the state of Washington. 
Participants in the study met the following inclusion criteria: 1) aged $18$ - $65$ years; 2) self-reported consumption of $4$ or more standard drinks on $5$ or more occasions in the past $30$ days; and 3) Diagnostic and Statistical Manual of Mental Disorders (DSM-$5$) diagnosis of moderate to severe AUD. Participants were also required to own a smartphone with a data plan that allowed them to respond to EMA surveys. Out of $20$ individuals screened, a total of $12$ agreed to participate, of which $11$ completed the study ($2$ did not meet eligibility, $6$ declined, and $1$ dropped out). On average, the participants were involved with the study for $14$ days with the range of $7$ - $17$ days. After written informed consent was obtained, participants were trained to use the E4 wristband during the first in-person meeting. During this initial session, participants were instructed to remove the wristband each night while sleeping and at other times during which the device may be damaged (e.g., in shower or bath) and to wear the device on the same wrist throughout the study, beginning the following day. The initial training session also included instructions about how to use the push-button interface on the E4 wristband that allows for real-time data annotation. Training for the EMA component occurred during the participants’ second in-person meeting, which occurred two days following the initial session. EMA data collection began the following morning. Moments of heightened stress were identified in two ways: 1) participants were asked to press the ``event marker button" on the E4 any time they felt``\textit{more stressed, overwhelmed, or anxious than usual}" and 2) EMA surveys were sent out $4$ times each day at set times included a question that asked ``\textit{Since [last assessment], have you experienced a time when you felt stressed, overwhelmed, or anxious than usual?}".

\begin{figure}[!tbh]
    \centering
    \includegraphics[width=0.9\linewidth]{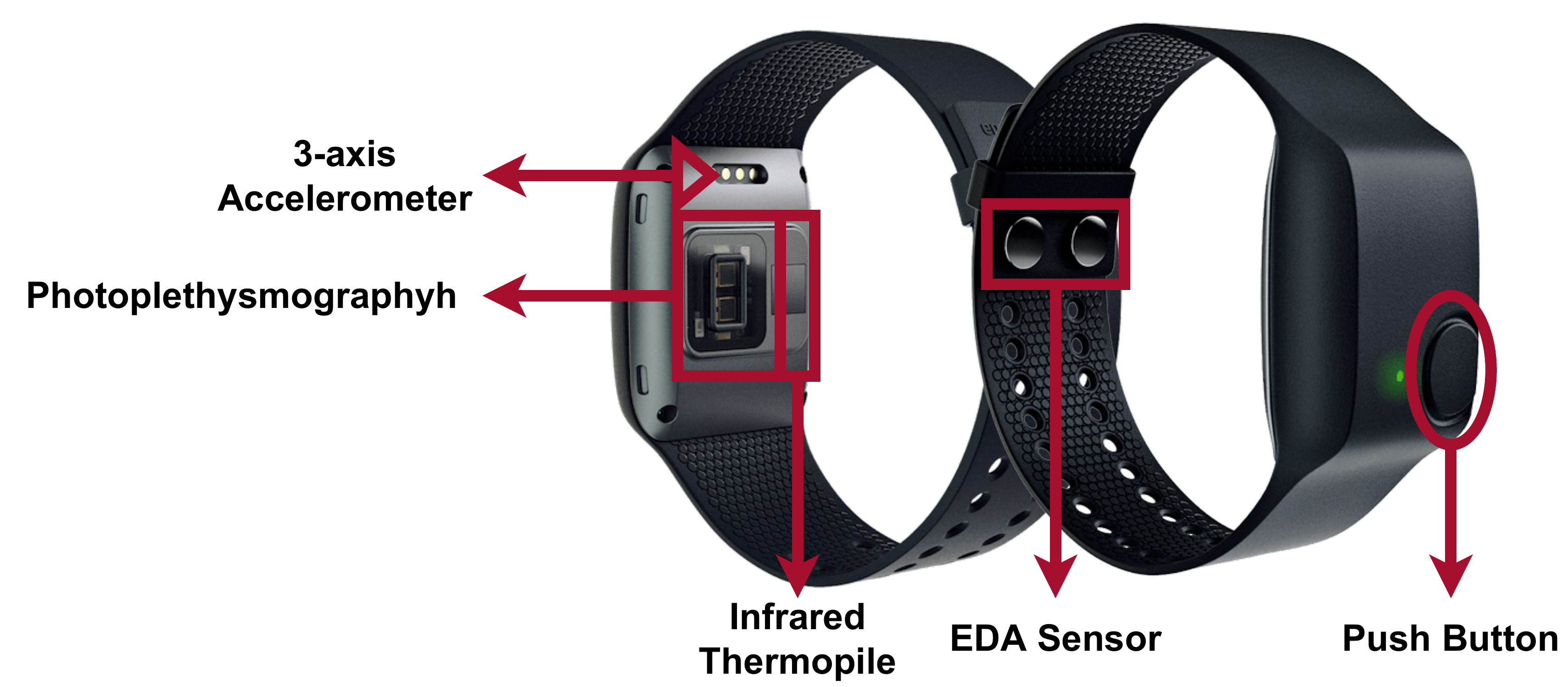}
    \caption{An Empatica E4 wristband with different sensors.}
    \label{fig:e4}
\end{figure}

\vspace{-5mm}

\subsection{Sensor System}
Empatica E4 is a wearable device that offers real-time continuous data acquisition of biomarkers associated with the arousal of the ANS due to stress. The device contains $4$ sensors: 1) photoplethysmography to provide Blood Volume Pulse (BVP), from which heart rate (HR), heart rate variability (HRV), inter-beat interval (IBI), and other cardiovascular features may be derived; 2) skin conductance or EDA, used to measure sympathetic nervous system arousal and to derive features related to stress, engagement, and excitement; 3) a $3$-axis accelerometer to capture motion-based activity; and 4) an infrared thermopile, used to measure skin temperature. The sampling frequency of the EDA sensor is $4$Hz with the resolution of $900$ pico Siemens and ranges between $0.01$ to $100 \mu S$. The accelerometer is sampled at $32$Hz and the infrared thermopile at $4$Hz. The E4 also includes a push-button for data annotation, as shown in \figref{e4}.

Previous research has assessed the validity of physiological signals recorded by an Empatica E4, such as EDA, HRV, and IBI, against the standard clinical ground truth \cite{schuurmans2020validity, milstein2020validating}. These studies indicate that E4 is among the most commonly used sensor devices in scientific research and validated its usefulness in detecting atrial fibrillation \cite{corino2017detection} and emotional arousal and stress \cite{ollander2016comparison, matsubara2016emotional}. 

\subsection{Ecological Momentary Assessment \& Interviews}
Participants responded to EMA surveys on their phones via a web browser. The surveys were sent $4$ daily and corresponded to early morning (waking), noon, late afternoon, and bedtime for the course of participation. The survey assessed the participants' perceptions of positive and negative emotions, alcohol-related carvings, and experiences of pain and discomfort. Positive and negative emotions were assessed using items drawn from the extended version of the Positive and Negative Affect Scale (PANAS) \cite{watson1997measurement}. Negative affect was assessed by asking, “[Since last assessment], have you felt [irritable / lonely / sad / guilty / ashamed / anxious / stressed]?” Similarly, participants reported their positive affect using the following terms: warmhearted, enthusiastic, affectionate, relaxed, calm, happy, joyful, and loving. Responses for each item were assessed on a 5-point scale ranging from 1 = not at all to 5 = extremely. Aggregate measures of daily negative and positive affect were created by first averaging across the respective items at each of the four assessments. 

The measurement of alcohol-related cravings was derived from previous research studies \cite{huhn2016ecological, tiffany1993development} and included three items: 1) “[since last assessment], the idea of using alcohol has intruded upon my thoughts”; 2) “[since last assessment], I have missed the feeling alcohol can give me”; 3) “[since last assessment], I have thought about how satisfying alcohol can be.” The response options included a $5 -$ point scale, ranging from $1$ (strongly disagree) to $5$ (strongly agree). Finally, pain and discomfort were measured with $2$ questions: 1) “[since last assessment], have you felt any physical discomfort?” and 2) “[since last assessment], have you felt any physical pain?” Five response options included: $1 = $ nonexistent, $2 = $ slight, $3 = $ moderate, $4 = $ intense, and $5 = $ unbearable. During the study, participants attended up to $6$ follow-up sessions to meet with study staff on an every-other-day basis. During the follow-up visits, the staff ascertained whether the participants were experiencing any difficulty and measured recent alcohol use using a chart of the US Standard Drink definition. At the end of participation, a final interview was conducted to assess significant events during the study and ascertain usability and comfort issues associated with using E4 wristbands in daily living conditions.



\subsection{Collected Data}
\begin{figure}[!tbh]
    \centering
    \includegraphics[width=\linewidth]{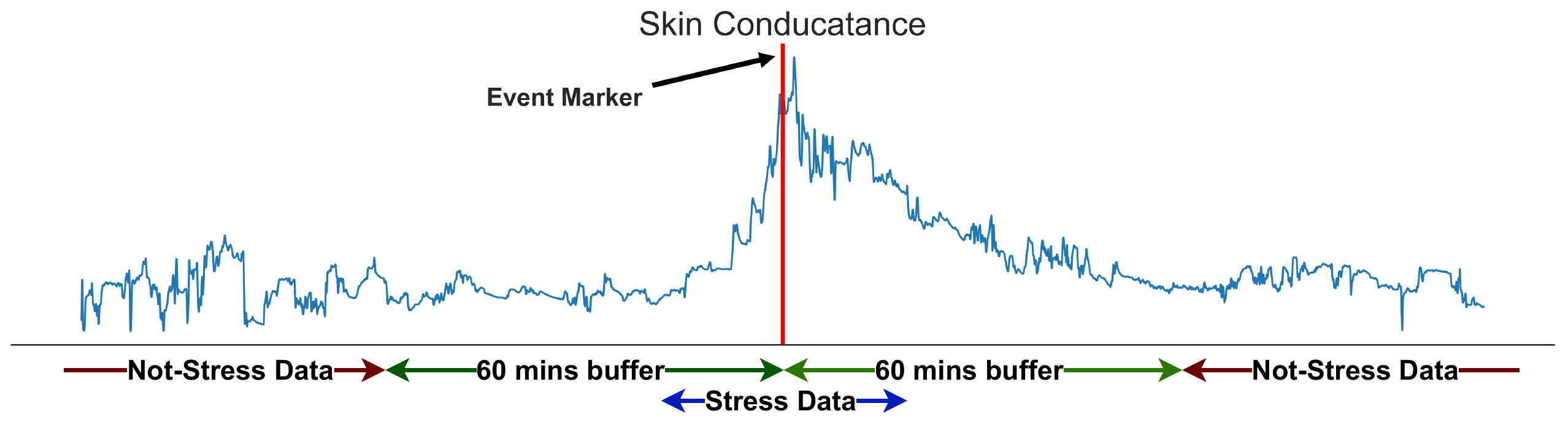}
    \caption{Partition of the sensor data into stress and not-stress class around a event marker.}
    \label{fig:single_event}
\end{figure}

A total of $1698$ hours of physiological sensor data were collected from $11$ participants, with $11.5$ hours of recording each day. The participants tagged $409$ events as moments of stress using the button available on the E4 wristband. On average, the participants tagged $37.2$ events with the range of $19$ - $87$ events. These events are points in the time indicated by the participants when they felt ``more stressed, overwhelmed, or anxious". We can visualize these events by plotting them over the sensor recording, for example, EDA as shown in \figref{single_event}.

\begin{figure}[!tbh]
    \centering
    \includegraphics[width=0.97\linewidth]{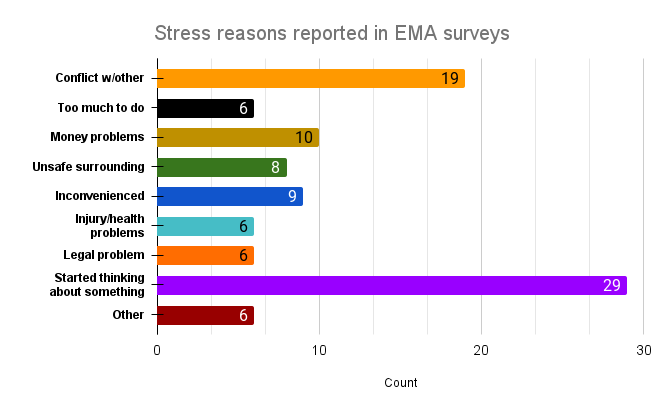}
    \caption{The number of times each reason for stress was selected by the participants in EMA surveys.}
    \label{fig:reason_stress}
\end{figure}

$10$ participants completed a total of $343$ surveys ($96$ morning, $90$ noon, $76$ late afternoon, and $81$ bedtime). Perceived stress events were reported in $99$ ($29.6\%$) of these surveys. On average, participants completed $34$ surveys (range $24$ - $39$) for a mean completion rate of $81.2$ (range $54.6$ - $100$). The mean number of stress events reported in the EMA surveys was $9.9$ (range $1$ - $24$). On average, participants reported stress events in $33.3\%$ of assessments (range $2.5\%$ - $96\%$). On surveys marked with stress events, morning surveys were most likely to be marked with stress events ($33$ out of $96$), and afternoon surveys had the least stress events ($20$ out of $76$). Participants also indicated the reason(s) for feeling stressed, overwhelmed, or anxious among the $9$ options for each event. \figref{reason_stress} highlights the number of times the participants selected each reason. 


\section{Results}
We present our initial analyses of the data from our study, including statistical analysis and machine learning algorithm design.

\subsection{Statistical Analysis}

We have demonstrated the relationship between physiological data and self-reported outcomes such as positive and negative emotions, experienced pain, and experienced discomfort \cite{alinia2021associations}. To this end, we first validated the quality of physiological signals collected via the E4 device from our user study. We determined that $87.9\%$ of the EDA signals were clean (i.e., valid data). Further, our preliminary results indicated that statistical features of both the EDA and HRV measures were significantly correlated with several self-reported outcomes provided by the other two study components (EMA surveys and qualitative interviews) at the person-level. Notably, EDA features, such as the amplitude of the signal and area under the curve, were moderately associated with the number of stressful events marked on the sensor device as well as self-reported positive and negative emotions, experienced pain, and experienced discomfort. Similarly, features of HRV, such as the mean of RR intervals (MRR) and mean of heart rate (MHR), were also significantly associated with participants’ reports of stressful events, positive and negative emotions, and experiences of pain and discomfort \cite{alinia2021associations}. 

\subsection{Stress Classification}

\begin{figure}[!tbh]
\includegraphics[width=\linewidth]{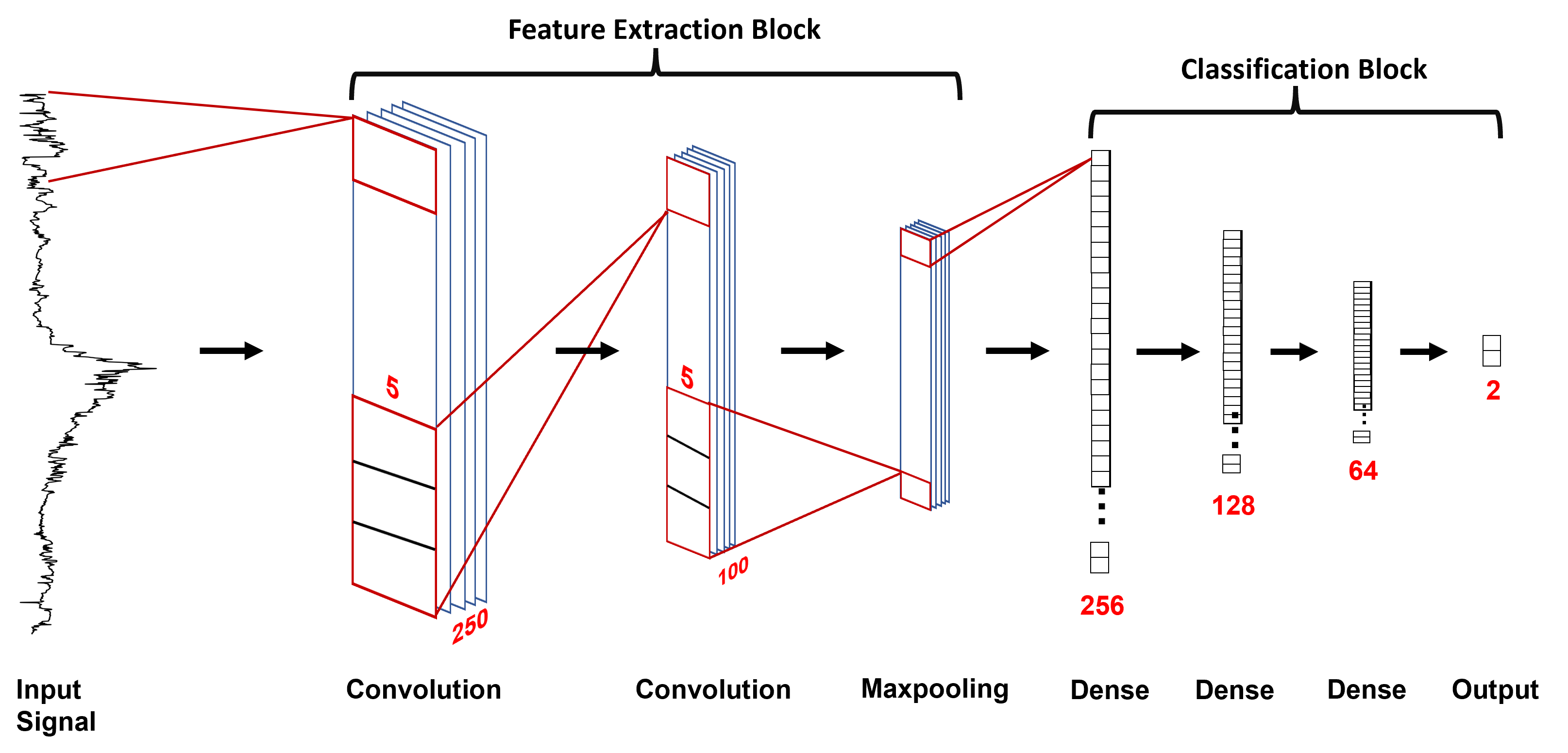}
  \caption{Architecture of 1D Convolutional Neural Network (CNN) model.}
  \label{fig:cnn_architecture}
\end{figure}

For training stress classification models with the ADARP dataset, we have used a Convolutional Neural Network (CNN) architecture. A CNN architecture allows us to use raw sensor data without cumbersome and expensive feature computation and selection process requiring domain knowledge. The CNN model comprises two $1D$ convolutional layers with $250$ and $100$ filters each and a kernel size of $5$. Convolutional layers are followed by a global max-pooling layer and three fully connected layers with $256$, $128$, and $64$ neurons. After the first and second fully connected layers, we have drop-out layers with a drop-out rate of $0.1$. The output layer has Softmax activation, and all other layers have ReLU activation as shown in \figref{cnn_architecture}. The learning rate was set to $0.001$, and categorical cross-entropy loss with Adam \cite{kingma2014adam} optimizer was used for training. The CNN model was trained for $50$ epochs with a batch size of $32$.




To train the CNN model, we only use the EDA sensor data and event markers obtained from the wristband. In total, $409$ timestamps were generated using the push button on the E4, and we labeled the sensor data around the generated timestamps as belonging to the stress class. \figref{single_event} shows the partition of the sensor data into the stress and not-stress class for a timestamp or event marker. Around each event marker, we have a buffer zone of $60$ minutes before and after the event. The data for stress class lies within this buffer region around the event marker, and all the data after the buffer zone is considered the not-stress class. If an event does not have these properties, it is dropped, and consequently, we have $181$ valid stress events. After removing the noise using a second-order Butterworth low-pass filter with a cut-off frequency of $1.25$ Hz, the EDA sensor data is normalized in the range $[0, 1]$. After normalization, we divide the stressed and not-stressed segments into $60$ seconds overlapping windows with $50\%$ overlap between consecutive windows. We settled on the window size of $60$ seconds because of available literature that has also used $60$ seconds window size for stress classification \cite{bobade2020stress, sah2020poster}.

\vspace{-5mm}

\begin{table}[tbh]
    \caption{Stress classification with majority class undersampling.}
    \label{tab:stress_classification_undersampling}
    \begin{tabular}{cccccc}
        \toprule
        {Dataset} & {Loss} & {Accuracy $(\%)$} & {Precision} & {Recall} & {f1-Score}\\
        \midrule
        Training & 0.0 & 99.72 & 0.99 & 0.99 & 0.99 \\
        Testing & 0.08 & 98.33 & 0.98 & 0.98 & 0.98\\
        \bottomrule
    \end{tabular}
\end{table}

\vspace{-5mm}

\begin{table}[tbh]
    \caption{Stress classification with minority class oversampling.}
    \label{tab:stress_classification_oversampling}
    \begin{tabular}{cccccc}
        \toprule
        {Dataset} & {Loss} & {Accuracy $(\%)$} & {Precision} & {Recall} & {f1-Score}\\
        \midrule
        Training & 0.0 & 99.92 & 0.99 & 0.99 & 0.99 \\
        Testing & 3.6 & 86.77 & 0.99 & 0.74 & 0.84 \\
        \bottomrule
    \end{tabular}
\end{table}

We get $163884$ samples for the not-stress class and $181$ for the stress class. The class imbalance observed in the ADARP dataset is one of the consequences of collecting data in real-world settings where stressors stimuli cannot be manufactured to the desired count compared to data collection in a lab setting with provisions for explicit stimuli to induce stress in participants. To balance the classes, we have used two different methods: 1) Minority Class Oversampling and 2) Majority Class Undersampling. In minority class oversampling, artificial data is generated for the minority class, and in majority class undersampling, samples from the majority class are randomly discarded to obtain an equal number of samples in all classes. For minority class oversampling, we use the Synthetic Minority Over-Sampling Technique (SMOTE) \cite{chawla2002smote} method to generate synthetic samples for the stress class. Tables \ref{tab:stress_classification_undersampling} and \ref{tab:stress_classification_oversampling} shows the performance of the trained model on the training and test sets. The model trained with majority class undersampling achieves state-of-the-art stress classification results with an f1-score of $0.99$. For minority class oversampling, the model's performance is lower on the test set compared to the training set.


\section{Conclusion}
We presented our study in detail and highlighted results from our analysis using the data we collected in real-world settings. The sensor and EMA survey data are made public to facilitate further research\footnote{\textit{https://zenodo.org/record/6640290}}. We also share code\footnote{\textit{https://github.com/rameshKrSah/ADARP\_Dataset}} that can be used to process the sensor data and extract meaningful information .

\vspace{-3mm}
\bibliographystyle{IEEEtran}
\bibliography{IEEEfull}

\begin{thebibliography}{10}
\providecommand{\url}[1]{#1}
\csname url@samestyle\endcsname
\providecommand{\newblock}{\relax}
\providecommand{\bibinfo}[2]{#2}
\providecommand{\BIBentrySTDinterwordspacing}{\spaceskip=0pt\relax}
\providecommand{\BIBentryALTinterwordstretchfactor}{4}
\providecommand{\BIBentryALTinterwordspacing}{\spaceskip=\fontdimen2\font plus
\BIBentryALTinterwordstretchfactor\fontdimen3\font minus
  \fontdimen4\font\relax}
\providecommand{\BIBforeignlanguage}[2]{{%
\expandafter\ifx\csname l@#1\endcsname\relax
\typeout{** WARNING: IEEEtran.bst: No hyphenation pattern has been}%
\typeout{** loaded for the language `#1'. Using the pattern for}%
\typeout{** the default language instead.}%
\else
\language=\csname l@#1\endcsname
\fi
#2}}
\providecommand{\BIBdecl}{\relax}
\BIBdecl

\bibitem{kasl}
\BIBentryALTinterwordspacing
S.~V. Kasl, ``Stress and health,'' \emph{Annual Review of Public Health},
  vol.~5, no.~1, pp. 319--341, 1984, pMID: 6372813. [Online]. Available:
  \url{https://doi.org/10.1146/annurev.pu.05.050184.001535}
\BIBentrySTDinterwordspacing

\bibitem{critchley2002electrodermal}
H.~D. Critchley, ``Electrodermal responses: what happens in the brain,''
  \emph{The Neuroscientist}, vol.~8, no.~2, pp. 132--142, 2002.

\bibitem{smets2018into}
E.~Smets, W.~De~Raedt, and C.~Van~Hoof, ``Into the wild: the challenges of
  physiological stress detection in laboratory and ambulatory settings,''
  \emph{IEEE journal of biomedical and health informatics}, vol.~23, no.~2, pp.
  463--473, 2018.

\bibitem{schuurmans2020validity}
A.~A. Schuurmans, P.~de~Looff, K.~S. Nijhof, C.~Rosada, R.~H. Scholte,
  A.~Popma, and R.~Otten, ``Validity of the empatica e4 wristband to measure
  heart rate variability (hrv) parameters: A comparison to electrocardiography
  (ecg),'' \emph{Journal of medical systems}, vol.~44, no.~11, pp. 1--11, 2020.

\bibitem{milstein2020validating}
N.~Milstein and I.~Gordon, ``Validating measures of electrodermal activity and
  heart rate variability derived from the empatica e4 utilized in research
  settings that involve interactive dyadic states,'' \emph{Frontiers in
  Behavioral Neuroscience}, p. 148, 2020.

\bibitem{corino2017detection}
V.~D. Corino, R.~Laureanti, L.~Ferranti, G.~Scarpini, F.~Lombardi, and L.~T.
  Mainardi, ``Detection of atrial fibrillation episodes using a wristband
  device,'' \emph{Physiological measurement}, vol.~38, no.~5, p. 787, 2017.

\bibitem{ollander2016comparison}
S.~Ollander, C.~Godin, A.~Campagne, and S.~Charbonnier, ``A comparison of
  wearable and stationary sensors for stress detection,'' in \emph{2016 IEEE
  International Conference on systems, man, and Cybernetics (SMC)}.\hskip 1em
  plus 0.5em minus 0.4em\relax IEEE, 2016, pp. 004\,362--004\,366.

\bibitem{matsubara2016emotional}
M.~Matsubara, O.~Augereau, C.~L. Sanches, and K.~Kise, ``Emotional arousal
  estimation while reading comics based on physiological signal analysis,'' in
  \emph{Proceedings of the 1st International Workshop on coMics ANalysis,
  Processing and Understanding}, 2016, pp. 1--4.

\bibitem{watson1997measurement}
D.~Watson and L.~A. Clark, ``Measurement and mismeasurement of mood: Recurrent
  and emergent issues,'' \emph{Journal of personality assessment}, vol.~68,
  no.~2, pp. 267--296, 1997.

\bibitem{huhn2016ecological}
A.~S. Huhn, J.~Harris, H.~H. Cleveland, D.~M. Lydon, D.~Stankoski, M.~J.
  Cleveland, E.~Deneke, and S.~C. Bunce, ``Ecological momentary assessment of
  affect and craving in patients in treatment for prescription opioid
  dependence,'' \emph{Brain research bulletin}, vol. 123, pp. 94--101, 2016.

\bibitem{tiffany1993development}
S.~T. Tiffany, E.~Singleton, C.~A. Haertzen, and J.~E. Henningfield, ``The
  development of a cocaine craving questionnaire,'' \emph{Drug and alcohol
  dependence}, vol.~34, no.~1, pp. 19--28, 1993.

\bibitem{alinia2021associations}
P.~Alinia, R.~K. Sah, M.~McDonell, P.~Pendry, S.~Parent, H.~Ghasemzadeh, M.~J.
  Cleveland \emph{et~al.}, ``Associations between physiological signals
  captured using wearable sensors and self-reported outcomes among adults in
  alcohol use disorder recovery: Development and usability study,'' \emph{JMIR
  Formative Research}, vol.~5, no.~7, p. e27891, 2021.

\bibitem{kingma2014adam}
D.~P. Kingma and J.~Ba, ``Adam: A method for stochastic optimization,''
  \emph{arXiv preprint arXiv:1412.6980}, 2014.

\bibitem{bobade2020stress}
P.~Bobade and M.~Vani, ``Stress detection with machine learning and deep
  learning using multimodal physiological data,'' in \emph{2020 Second
  International Conference on Inventive Research in Computing Applications
  (ICIRCA)}.\hskip 1em plus 0.5em minus 0.4em\relax IEEE, 2020, pp. 51--57.

\bibitem{sah2020poster}
R.~K. Sah, H.~Ghasemzadeh, A.~Habibi, M.~McDonell, P.~Patricia, and
  M.~Cleveland, ``Poster: Mobile health for alcohol recovery and relapse,'' in
  \emph{2020 IEEE/ACM International Conference on Connected Health:
  Applications, Systems and Engineering Technologies (CHASE)}.\hskip 1em plus
  0.5em minus 0.4em\relax IEEE, 2020, pp. 18--19.

\bibitem{chawla2002smote}
N.~V. Chawla, K.~W. Bowyer, L.~O. Hall, and W.~P. Kegelmeyer, ``Smote:
  synthetic minority over-sampling technique,'' \emph{Journal of artificial
  intelligence research}, vol.~16, pp. 321--357, 2002.

\end{thebibliography}

\end{document}